\documentclass[twocolumn,superscriptaddress,showpacs]{revtex4-2}
\usepackage{color}
\definecolor{red}{rgb}{0.75,0,0}
\definecolor{blue}{rgb}{0,0,0.75}
\definecolor{green}{rgb}{0,0.5,0}

\usepackage{amsmath}
\usepackage{amssymb}
\usepackage{lipsum}
\usepackage{bm}
\usepackage{xcolor}
\usepackage{CJK}
\usepackage{dcolumn}
\usepackage{graphicx}
\usepackage{verbatim}
\usepackage[T1]{fontenc}
\usepackage{hyperref}
\usepackage[normalem]{ulem}
\usepackage{datetime}
\longdate
\usepackage{epstopdf}
\makeatletter
\def\maketitle{
	\@author@finish
	\title@column\titleblock@produce
	\suppressfloats[t]}
\makeatother
\usepackage{mathrsfs}  
\usepackage[T1]{fontenc}

\def\bea{\begin{eqnarray}}
\def\eea{\end{eqnarray}}

\def\D2{\boldsymbol{\mathcal{D}}}
\def\besub{\begin{subequations}}
\def\eesub{\end{subequations}}

\def\bwd{\begin{widetext}}
\def\ewd{\end{widetext}}

\definecolor{ao(english)}{rgb}{1.0, 0.0, 0.0}
\definecolor{armygreen}{rgb}{0.29, 0.33, 0.13}
\definecolor{auburn}{rgb}{0.43, 0.21, 0.1}
\definecolor{brightmaroon}{rgb}{0.76, 0.13, 0.28}
\definecolor{cadmiumred}{rgb}{0.89, 0.0, 0.13}
\definecolor{carnelian}{rgb}{0.7, 0.11, 0.11}
\definecolor{cornellred}{rgb}{0.7, 0.11, 0.11}
\definecolor{crimsonglory}{rgb}{0.75, 0.0, 0.2}
\definecolor{orangeyellow}{rgb}{0.3, 0.2, 0.2}
\definecolor{fluorescentorange}{rgb}{1.0, 0.75, 0.0}
\definecolor{gamboge}{rgb}{0.89, 0.61, 0.06}
\newcommand{\bsf}[1]{\textsf{\textbf{#1}}}

\newenvironment{frcseries}{\fontfamily{frc}\selectfont}{}

\begin{document}

\title{Dynamics of Ordered Active Columns: Flows, Twists, and Waves}

\author{S. J. Kole}
\email{swapnilkole@iisc.ac.in}
\affiliation{Centre for Condensed Matter Theory, Department of Physics, Indian Institute of Science, Bangalore 560 012, India}

\author{Gareth P. Alexander}
\email{G.P.Alexander@warwick.ac.uk}
\affiliation{Department of Physics, Gibbet Hill Road, University of Warwick, Coventry CV4 7AL, United Kingdom}

\author{Ananyo Maitra}
\email{nyomaitra07@gmail.com}
\affiliation{{Laboratoire de Physique Th\'eorique et Mod\'elisation, CNRS UMR 8089,
		CY Cergy Paris Universit\'e, F-95032 Cergy-Pontoise Cedex, France}}
\affiliation{Sorbonne Universit\'{e} and CNRS, Laboratoire Jean Perrin, F-75005, Paris, France}

\author{Sriram Ramaswamy}
\email{sriram@iisc.ac.in}
\affiliation{Centre for Condensed Matter Theory, Department of Physics, Indian Institute of Science, Bangalore 560 012, India}


\begin{abstract}
We formulate the hydrodynamics of active columnar phases, with two-dimensional translational order in the plane perpendicular to the columns and no elastic restoring force for relative sliding of the columns, using the general formalism of an active model H$^*$. Our predictions include: two-dimensional odd elasticity coming from three-dimensional plasmon-like oscillations of the columns in chiral polar phases with a frequency that is independent of wavenumber and non-analytic; a buckling instability coming from the generic force-dipole active stress analogous to the mechanical Helfrich-Hurault instability in passive materials; the selection of helical column undulations by apolar chiral activity. 
\end{abstract}

\maketitle

Living materials form a plethora of organised structures. Whereas many of these have the spatial asymmetries seen in ordered phases at thermal equilibrium, living matter is ``active'', continually converting chemical energy into work. A description of the mechanics and statistics of active systems must either explicitly account for this chemistry or introduce a parameter that breaks time-reversal symmetry at the microscopic level \cite{SRJSTAT, LPDJSTAT, RMP, SRrev, Prost_nat, Salbreux_PhysRep}. The novel features of the resulting dynamics have been extensively examined for orientationally ordered phases, but translational order has received less attention \cite{Tap_smec,Tap_chol, julicher2022broken,Ano_sol, adar_joanny, Ano_hex, Deboo2,SJ_chiral_layered}. In particular, despite the abundance of filamentous assemblies in soft matter and biology \cite{barberi2021local,Lubke,grason2015colloquium,atkinson2021mechanics}, the active hydrodynamics of a two-dimensional crystalline arrangement of fluid columns \cite{chandrasekhar1977liquid,prost1980liquid,ramaswamy1983breakdown,deGen} in three dimensions has not been explored. Apart from the breaking of continuous rotational and translational invariance, two discrete asymmetries play a central role in the present work. \textit{Chiral} symmetry \cite{Lubensky_Kamien} is broken when the elements composing the columns lack mirror symmetry. \textit{Polar} asymmetry implies that the columnar material lacks inversion symmetry along the fluid direction -- i.e., symmetry under the transformation $z\to -z$ where $\hat{\bf z}$ is the mean direction of the column tangents, orthogonal to the crystalline $\perp \equiv xy$ plane. Both of these asymmetries are widely present in living and artificial columnar systems \cite{polar_columnar_exp}. The most natural example of active columnar materials are axons, whose subunits are both chiral and polar. Indeed, the microtubule bundles comprising the axon in nerve cells are actively polarity-sorted and therefore possess macroscopic polar order and chirality \cite{rao2018polarity,ahmadi2006hydrodynamics}. 
In this article, we will therefore consider four cases: apolar and achiral, polar and achiral, apolar and chiral, and polar and chiral. We assume chirality leaves the columnar structure intact and does not supplant it with an analogue of the TGB phases \cite{TGB_columnar}.

\begin{figure*}[t]
\centering
\includegraphics[width=1.0\linewidth]{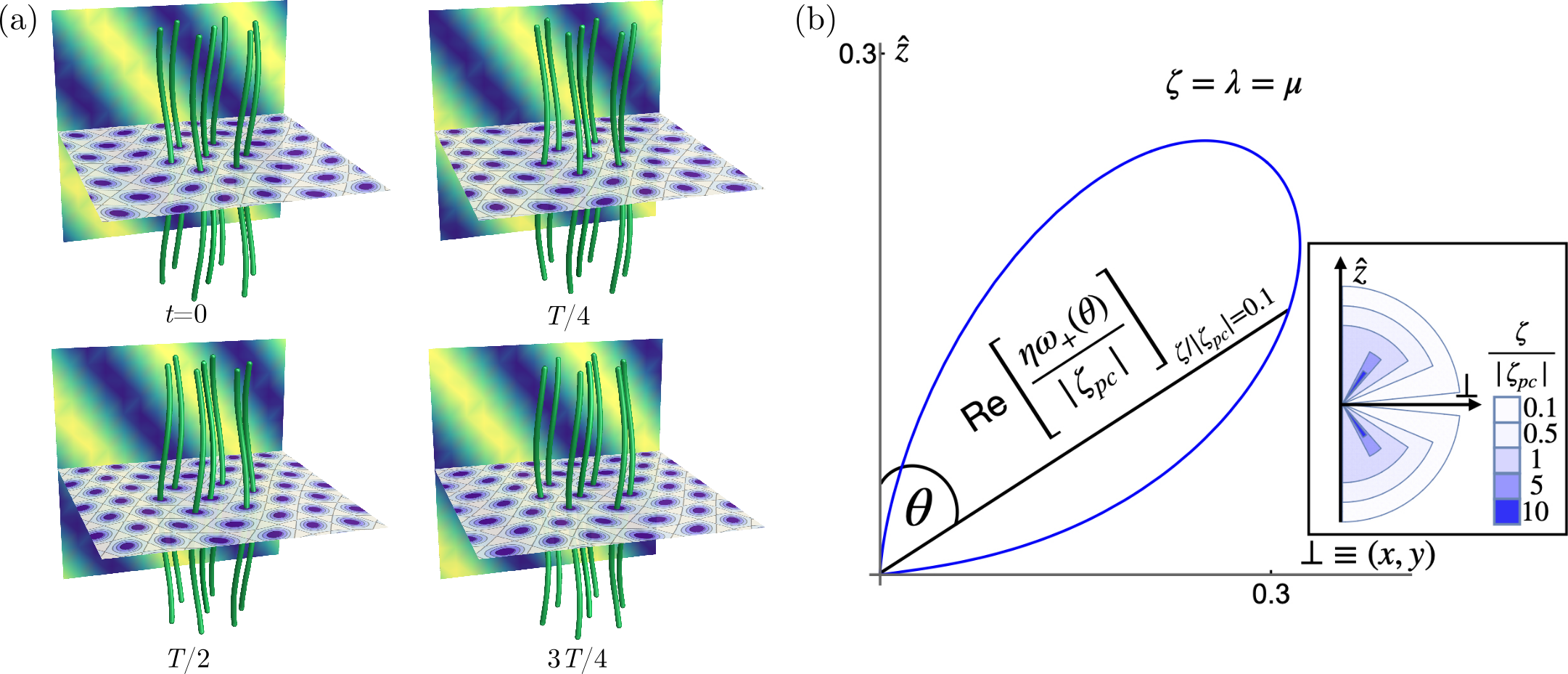}
\caption{(a) Column oscillations of an active polar, chiral phase for a plane wave solution of~\eqref{ulpol} and~\eqref{utpol}. The horizontal plane shows a density plot of $\psi$ and the background plane indicates the phase and direction of the plane wave. Images are shown for every quarter of an oscillation period. (b) {Polar plot of $\text{Re}[\eta\omega(\theta)_+/|\zeta_{pc}|]$---the oscillation frequency, non-dimensionalised by the active timescale $\eta/|\zeta_{pc}|$---of the odd-elastic plasmon-like mode in the polar chiral columnar phase, for $\lambda=\mu=\zeta$. $\theta=0$ corresponds to a perturbation purely along the $\hat{{\bf z}}$ (polarity) direction while $\theta=\pi/2$ corresponds to a perturbation purely along the planar, crystalline directions. \textit{Inset}: Colour-coded sectors (with different radii to ensure visibility) showing the range of $\theta$---the angle between the wavevector of perturbation and $\hat{{\bf z}}$---that elicits an oscillatory response for various values of $\zeta/|\zeta_{pc}|$. This demonstrates that the angular range increases with decreasing $\zeta/|\zeta_{pc}|$. 
While the angular range for $\zeta/|\zeta_{pc}|\leq1$ extends \emph{almost} to $\theta=0$, there is no oscillatory response for a perturbation with wavevector along $\hat{{\bf z}}$. Instead, $\text{Re}[\omega_\pm]\propto\theta^2$ at small $\theta$.}}
\label{fig:polar_wave_3d}
\end{figure*}
Our most interesting results concern polar, chiral columnar materials. We find that such three-dimensional liquid crystals display odd \emph{two-dimensional} translational elasticity, with the two components of the displacement field mimicking a position-momentum pair \cite{Ano_hex, Deboo2}. Unlike other examples of planar odd elasticity, for instance, in materials composed of two-dimensional spinners \cite{tan2022odd} or two-dimensional crystals with ``odd springs'', here odd elasticity emerges naturally from the columnar organisation of three-dimensional chiral active units. 
When the odd elastic modulus $\zeta_{pc}$ is large enough, the interplay of oddness-induced non-reciprocity and Stokesian hydrodynamics with viscosity $\eta$ leads to an \emph{oscillatory, plasmon-like mode} for perturbations along most  directions of the wavevector ${\bf q} \equiv ({\bf q}_\perp, q_z)$. The existence and frequency of this mode depends on the angle between ${\bf q}$ and $\hat{\bf z}$ but, crucially, not on the wave\textit{number}. The squared vibrational frequency is $(\zeta_{pc}/\eta)^2 q_\perp^4 q_z^2/q^6$. The three-dimensional character of the columnar material is essential for this oscillation: the Poisson-bracket-like coupling between the displacement fields vanishes for distortions purely along the column (fluid) direction or purely transverse to it (i.e., purely in the crystalline directions) and, therefore, there is no oscillatory mode for such perturbations. With Lam\'e coefficients $\lambda$ and $\mu$ of the two-dimensional elasticity, and an active achiral stress $\zeta$, Fig.~\ref{fig:polar_wave_3d}{(a) displays the stroboscopic, three-dimensional conformations of the columns every quarter of an oscillation period and (c) shows a polar plot of the frequency of the oscillatory mode -- non-dimensionalised by the active frequency $|\zeta_{pc}|/\eta$ -- as a function of the angle $\theta$ from $\hat{{\bf z}}$. We also display the range of wavevector angles that elicit an oscillatory response for various values of parameters in the inset of Fig.~\ref{fig:polar_wave_3d}(b).}

Another key result is an active column tension, positive or negative depending on whether the uniaxial active achiral stress is contractile or extensile. Contractile systems thus have displacement fluctuations with wavenumber $q$ have variance of order $1/q^2$ in all directions, while extensile systems display a spontaneous column-undulation instability.  We demonstrate that the active achiral stress in columnars is functionally equivalent to an externally imposed in-plane, isotropic stress, allowing us to calculate the extensile instability threshold by a mapping to the columnar Helfrich-Hurault effect.

While polar and apolar \textit{achiral} active columnar materials are hydrodynamically equivalent, chirality breaks this degeneracy. In place of the oscillatory modes predicted for polar systems, the apolar phases display active chiral fluid flows that deform the columns into helices.

We now show how we obtain these results. We start from the coupled dynamics of a conserved scalar field $\psi$ and the density ${\bf g} = \rho{\bf v}$ of a conserved momentum, with overall incompressibility: a constant total mass density $\rho$ and $\nabla \cdot {\bf v} = 0$ for the total velocity field ${\bf v}$. The equations of motion of the columnar phases emerge upon expanding about a state in which $\psi$ has the form of a two-dimensional density wave. We break detailed balance in Model H in the Hohenberg-Halperin \cite{HalpHohen} classification by introducing active stresses and currents~\cite{CatesH1,CatesH2, SJ_chiral_layered}. Of these the simplest is $\sigma^a_{ij}\propto\partial_i\psi\partial_j\psi$. If the material is chiral, contributions proportional to the Levi-Civita tensor arise, such as $\sigma^{ac}_{ij}\propto [\epsilon_{ilk}\partial_l(\partial_k\psi\partial_j\psi)]^S$, and $\sigma^{pc}_{ij}\propto[\epsilon_{ikl}P_l\partial_k\psi\partial_j\psi]^S$ if the material possesses a non-zero polar order parameter ${\bf P}$, where the superscript $S$ denotes symmetrisation. In systems with only orientational order, note that while $\sigma^{ac}_{ij}$ \cite{SJ_chiral_layered} has a nematic analogue $[\epsilon_{ilk}\partial_lQ_{kj}]^S$, where ${\bsf Q}$ is the nematic order parameter \cite{Cates_chi,Cates_drop, furthauer2012, Cates_Marko}, $\sigma^{pc}_{ij}$ has no polar analogue. 

In the supplement, we detail how we obtain the dynamical equations for active columnar fluids from these extended active model H equations. The broken-symmetry hydrodynamics of columnar materials requires the phases of the $\psi$-density wave, that is the Nambu-Goldstone modes of spontaneously broken translation symmetry, denoted by the two-component vector ${\bf u}_\perp({\bf r}, t)$, where ${\bf r}\equiv(x,y,z)$ is the three-dimensional position vector. 

Ignoring permeation, which we will show later to be irrelevant at long wavelengths, the linearised equation of motion for ${\bf u}_\perp$ is $\partial_t{\bf u}_\perp={\bf v}_\perp$. In the absence of activity, the dynamics would be controlled by a free energy  $F_u=(1/2)\int_{\bf r}[\lambda(\mathrm{Tr}{\bsf E})^2+2\mu{\bsf E}:{\bsf E}+K\nabla^2{\bf u}_\perp\cdot\nabla^2{\bf u}_\perp]$ \cite{deGen}, where $E_{ij}=(1/2)(\partial_i u_j+\partial_ju_i)$ is the strain, here written in a linearised approximation, leading to a linearised force density $-\delta F_u/\delta{\bf u}_\perp$. In active materials, extra non-equilibrium force densities arise that cannot be obtained from a functional derivative  of a rotation-invariant functional. In this article, we will examine the effects of three such force densities. The first, with a coefficient $\zeta$, is present in all active columnar materials; the second, with a coefficient $\zeta_c$, requires chiral asymmetry and the third, with a coefficient $\zeta_{pc}$, requires both polar and chiral asymmetries. They arise, respectively, from the stresses $\bm{\sigma}^a$, $\bm{\sigma}^{ac}$ and $\bm{\sigma}^{pc}$ in our active model H$^*$. Upon neglecting inertia, as appropriate for the slow flows arising in most soft-matter and biological contexts, the conservation of momentum reduces to balancing viscous force densities against active and passive force densities involving ${\bf u}_\perp$:
\begin{multline}
	\label{frcbal}
	\eta\nabla^2{\bf v} = \nabla\Pi - \mu\nabla_\perp^2{\bf u}_\perp - (\lambda+\mu)\nabla_\perp\nabla_\perp\cdot{\bf u}_\perp \\
    - \zeta\nabla^2{\bf u}_\perp - \zeta_c\nabla^2\nabla\times{\bf u}_\perp - \zeta_{pc}\nabla_\perp^2\bm{\epsilon}\cdot{\bf u}_\perp + \bm{\xi}_v,
\end{multline} 
where the pressure $\Pi$ enforces incompressibility, $\bm{\xi}_v$ is a zero-mean, Gaussian white noise whose correlator is $-2T\eta\nabla^2\delta({\bf r}-{\bf r}')\delta(t-t')$ and $\bm{\epsilon}$ is the rank-2 Levi-Civita tensor. For polar materials there is also an achiral active force density at $\mathcal{O}(\nabla^3)$ $\propto \nabla^2\partial_z{\bf u}_\perp$. We focus on the lower-order polar term but make some remarks about this contribution later on. To the lowest order in wavenumbers, the dynamics of \textit{achiral} polar and apolar active columnar materials are exactly equivalent.

We start by examining the response of achiral and apolar active columnar liquid crystal to a perturbation with wavevector ${\bf q}=q(\sin\theta\cos\phi,\sin\theta\sin\phi,\cos\theta)$.
Setting $\zeta_c=\zeta_{pc}=0$ in \eqref{frcbal}, projecting it transverse to the wavevector, eliminating the velocity field and defining longitudinal and transverse displacement fields $u_l={\bf q}_\perp\cdot{\bf u}_\perp/|q_\perp|$ and $u_t=(q_xu_y-q_yu_x)/|q_\perp|$, we obtain the decoupled dynamics which is independent of the wave\textit{number} $q$ for $q \to 0$:
\begin{equation}
	\label{achid1}
	\dot{u}_l=-\frac{\cos^2\theta}{\eta}\left[{(2\mu+\lambda)\sin^2\theta}+{\zeta}\right]u_l+\xi_l 
\end{equation}
and
\begin{equation}
		\label{achid2}
	\dot{u}_t=-\frac{1}{\eta}\left[{\mu\sin^2\theta}+{\zeta}\right]u_t+\xi_t. 
\end{equation}
The white noises $\xi_l$ and $\xi_t$ have variances $2T\cos^2\theta/\eta q^2$ and $2T/\eta q^2$, so that the correlators for $u_l$ and $u_t$ both scales as $\sim 1/q^2$ (for $\zeta>0$) along \emph{all} spatial directions {(thereby cutting off the viscosity divergences that arise in the equilibrium columnar phase \cite{ramaswamy1983breakdown}\footnote{{Similarly, viscosity divergences of equilibrium smectics \cite{Ramaswamy_Toner_Mazenko,Ramaswamy_Toner_Mazenko_PRL} are cut-off by the suppression of fluctuations in active layered materials \cite{Tap_smec, SJ_chiral_layered, Tap_chol}}})} despite the relaxation rates scaling as $\sim q^0$. This, however, crucially requires $\zeta>0$. In a passive columnar system $\zeta=0$, the displacement fluctuations would diverge as $\sim 1/q^4$ for fluctuations along the column direction. Activity leads to a column tension. However, when $\zeta<0$, the column tension is destabilising and destroys the columnar phase. 

The effect of the active apolar stress $\zeta$ in an incompressible columnar system is \emph{exactly} the same as that of an external stress, uniaxial along the columns or, equivalently, isotropic in the plane normal to them, as we show in the supplement. Thus, as in cholesterics and smectics \cite{SJ_chiral_layered}, $\zeta<0$ leads to a spontaneous Helfrich-Hurault instability with degeneracy in the polarisation of the column undulation, which is shown in Fig.~\ref{oddflows} for a planar undulation. For a finite system of size $[-L/2,L/2]^2\times(-\infty,\infty)$ with a helical modulation of the columns with displacement field ${\bf u}(x,y,z) = \alpha x \,{\bf e}_x + \alpha y \,{\bf e}_y +  u_0 \cos(\pi x/L) \cos(\pi y/L)\bigl[ \cos qz \,{\bf e}_x + \sin qz \,{\bf e}_y \bigr]$ the mode that goes unstable first is $q^2=({\pi}/{L})\sqrt{{\mu}/{K}}$ for a threshold strain of $\alpha=({\pi}/{L})\sqrt{{\mu}/{\lambda}}$. Equivalently, the critical active stress for a finite system is $\zeta_{th}=\pi{\sqrt{K\lambda}}/{L}$, details of the calculation can be found in the supplement \cite{supp}.

\begin{figure}[t]
	\centering
	\includegraphics[width=1\linewidth]{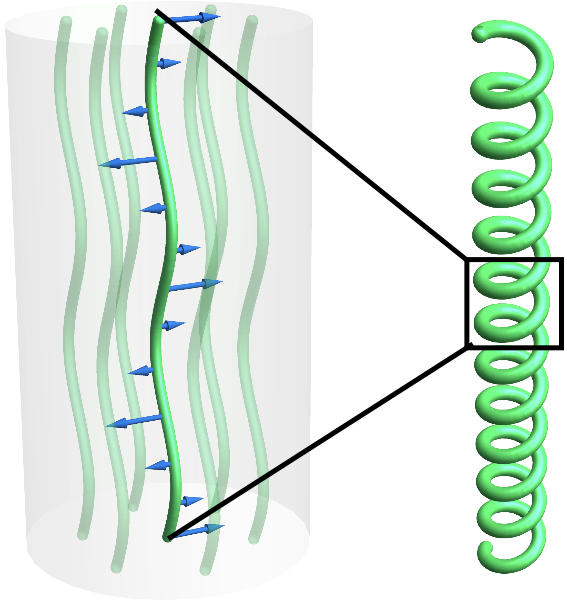}
	\caption{{The columnar phase suffers a spontaneous Helfrich-Hurault instability when $\zeta<0$ resulting in the buckling of columns. In chiral systems, $\zeta_c\neq 0$ and even planar distortions produced by this instability lead to flows (blue arrows) that tend to twist columns into helices.}}
	\label{oddflows}
\end{figure}

We now examine the chiral active force density $\propto \zeta_c$ in \eqref{frcbal}. Though sub-leading in gradients, this term is physically interesting as it couples longitudinal and transverse fluctuations:
\begin{equation}
	\dot{u}_l=-\frac{\cos^2\theta}{\eta}\left[{(2\mu+\lambda)\sin^2\theta}+{\zeta}\right]u_l-\frac{iq\zeta_c\cos\theta}{\eta}u_t+\xi_l,
\end{equation}
and
\begin{equation}
	\dot{u}_t=-\frac{1}{\eta}\left[{\mu\sin^2\theta}+{\zeta}\right]u_t+\frac{iq\zeta_c\cos\theta}{\eta}u_l
	+\xi_t, 
\end{equation}
yielding, for perturbations purely along the column direction, $\hat{{\bf z}}$, mode frequencies $\omega_\pm=-({i}/{\eta})[\zeta\pm\zeta_c q]$. That is, chiral activity reduces the relaxation rate of one of the modes while enhancing that of the other. Interestingly, unlike odd elasticity in two-dimensional chiral solids~\cite{Deboo2}, $\zeta_c$ here is a \emph{dissipative} Onsager-like coupling between $u_t$ and $u_l$. Note that this coupling originates from the same stress in active model H$^*$ that led to an odd elastic force density in active layered materials directed primarily along contours of constant mean curvature of the layers~\cite{SJ_chiral_layered}. For the columnar phase, in contrast, these transverse chiral flows generate a deformation of the columns into helices even at linear order (see Fig.~\ref{oddflows}). 

We now turn to phases with macroscopic polarity along $\hat{{\bf z}}$. As discussed earlier, polar and apolar achiral columnar materials are distinguishable only at sub-leading order in wavenumber. An \textit{achiral} but polar active force density $\zeta_p\partial_z\nabla^2{\bf u}_\perp$ leads to a simple modification of \eqref{achid1}, \eqref{achid2}: $\zeta\to \zeta+i\zeta_pq\cos\theta$. Chirality, by contrast, breaks the degeneracy of polar and apolar active columnar liquid crystals at \textit{leading} order in wavenumber. Intriguingly, the polar, chiral active force density $\zeta_{pc}\nabla_\perp^2\bm{\epsilon}\cdot{\bf u}_\perp$ in \eqref{frcbal} is exactly the odd elasticity discussed in \emph{two-dimensional} active solids \cite{Deboo2, Ano_hex}, but here realised in a fully three-dimensional active material. The resulting coupling of longitudinal and transverse displacements 
\begin{equation}
	\label{ulpol}
	\dot{u}_l = - {(2\mu+\lambda)q_{\perp}^2 q_z^2 + \zeta q^2 q_z^2 \over \eta q^4} u_l - {\zeta_{pc} q_z^2 q_{\perp}^2 \over \eta q^4} u_t + \xi_l
\end{equation}
and
\begin{equation}
	\label{utpol}
	\dot{u}_t = - {\mu q_{\perp}^2 +\zeta q^2 \over \eta q^2} u_t + {\zeta_{pc}q_{\perp}^2 \over \eta q^2} u_l + \xi_t,
\end{equation}
enters at zeroth order in wavenumber and has a Poisson-bracket-like structure. This coupling is truly three-dimensional, vanishing for $\theta = 0$ or $\pi/2$. Its equivalent is ruled out in an incompressible, two-dimensional odd poroelastic gel, where to $\mathcal{O}(q^0)$, $\dot{u}_l=v_l=0$. Here, however, incompressibility constrains the three-dimensional velocity field and, therefore, $v_l\neq 0$ for a perturbation in a general direction; of course, $\dot{u}_l$ vanishes (at this order in wavenumbers) for perturbation purely in the solid directions. As discussed in the introduction, \eqref{ulpol} and \eqref{utpol} can lead to an oscillatory response for perturbations with $\theta\neq0,\pi/2$ for large-enough $\zeta_{pc}$. While the general expression for the eigenfrequency displayed in \cite{supp} is complicated, for $\theta=\pi/4$ it assumes the form
\begin{equation}
	 	\label{osceigen}
	 	\omega_\pm(\theta=\pi/4) = \pm \frac{\mid{\zeta_{pc}}\mid}{2\sqrt{2}\eta}\sqrt{1 - \left(\frac{\lambda - 2\zeta}{2\sqrt{2}\zeta_{pc}}\right)^2} - i \frac{6\zeta+4\mu+\lambda}{8\eta}.
\end{equation} 
Notice that \eqref{osceigen} describes an overdamped mode when $|\zeta_{pc}| < |\lambda-2\zeta|/2\sqrt{2}$. In general, perturbations with wavevector making an angle $\theta\neq 0,\pi/2$ with $\hat{{\bf z}}$  will elicit an oscillatory response when $|\zeta_{pc}|>|\zeta+\mu-(2\mu+\lambda)\cos^2\theta|/2|\cos\theta|$. {This implies that when $\zeta_{pc}$ is larger than the other active and elastic terms, a perturbation with a wavevector \emph{almost} fully along $\hat{{\bf z}}$ also leads to an oscillatory response with $\lim_{\theta\to 0}\mathrm{Re}[\omega_\pm]\propto\theta^2$. We display the angular ranges in which we predict an oscillatory response in Fig. \ref{fig:polar_wave_3d}(b) for various values of $\zeta_{pc}$.}

An instructive comparison is with the dispersive wave induced by non-reciprocity in two-dimensional compressible solids on frictional substrates \cite{Deboo2}, where odd elasticity leads to the dynamics  $\dot{u}_l\propto -q_\perp^2 u_t$ and $\dot{u}_t\propto q_\perp^2 u_l$. In active polar and chiral columnar phases, the presence of momentum conservation and three-dimensional incompressibility leads to a radically different mode structure. Momentum conservation induces a long-range interaction, replacing the dispersive wave by a plasmon-like oscillatory mode that is non-analytic in the $q\to 0$ limit. In the presence of a momentum sink, the damping in the momentum equation would ultimately imply a dispersive wave at small wavenumbers. Indeed, if the columnar material is confined in a geometry which has a finite extent $\ell$ in the $z$ direction, with porous walls such that the fluid can flow out of the system, then one can replace $q_z^2$ with $1/\ell^2$ to obtain the displacement dynamics in \cite{Deboo2} quoted above. Note that the three-dimensional character of the active liquid crystal is still crucial for this: it requires a $z$-variation of the $z$-component of the velocity field at the scale of the system. 

If instead the confinement is on a scale $\ell$ in the $\perp$ plane, the odd elastic coupling for $q_z \ell \ll 1$ takes the form $\dot{u}_l\propto -(\zeta_{pc}/\eta)\ell^2 q_z^2u_t$ and $\dot{u}_t\propto (\zeta_{pc}/\eta) u_l$. The possibility of active non-dispersive waves \`{a} la \cite{RamTonPro, Ano_mem} is, however, ruled out by the effective damping rates $(\mu+\zeta)/\eta$ and $(2\mu+\lambda+\zeta)\ell^2q_z^2/\eta$ of $u_t$ and $u_l$ respectively. The resulting eigenfrequencies $\omega_+=-i(\zeta+\mu)/\eta$ and $\omega_-=-i\ell^2q_z^2[(\zeta+\lambda+2\mu)/\eta + \zeta_{pc}^2/\eta(\zeta+\mu)]$ have no real part. 

We have described the hydrodynamics of active columnar phases, including both chiral and polar materials. The polar chiral activity realises a form of two-dimensional odd elasticity, with a plasmon-like oscillatory response of the columns whose frequency scales as $q^0$ and is non-analytic as $q \to 0$. Three-dimensionality and viscous hydrodynamic interaction are essential to the mechanism. Active columnar phases are an ideal prospect for both natural and synthetic odd elastic materials. Biological tissues, such as axons and epithelia, have a columnar structure and a natural polarity, so chiral activity in these tissues should generate odd elastic effects. Columnar liquid crystals with macroscopic polarity \cite{polar_columnar_exp}, if suffused with chiral microswimmers, could realize a material in which to test our predictions. 

The achiral active stress is analogous to an applied mechanical stress and produces a Helfrich-Hurault instability of the columns with degeneracy in the polarisation of the undulations. As in active cholesterics, an imposed stress can be used to control the instability. The bend instability in three-dimensional extensile nematics \cite{Aditi1, RMP} is generically accompanied by twist to satisfy geometric compatibility \cite{pollard2021,Rayan,shendruk2018,baskaran_dogic}. Analogously, the active columnar instability could favour spontaneous chiral-symmetry breaking through a coherent twisting of the column undulations. In \textit{chiral} active columnar materials, $\zeta_c$ explicitly breaks parity, favouring one sign of helical column undulation. This highlights the emergence of qualitatively different effects from the same chiral stress in distinct broken-symmetry phases of active model H$^*$. 

An interesting extension of our results would be to determine the effects of confinement and boundary conditions on the column undulations and their associated odd elasticity. Finally, our predictions are based largely on linear stability analysis. An understanding of the evolution beyond the linear regime and the final state of the system requires a direct numerical solution of the nonlinear equations of motion. 

\acknowledgements{We thank Jacques Prost for valuable discussions. SJK acknowledges support through a Raman-Charpak Fellowship of CEFIPRA, hosted by Cesare Nardini, SPEC, CEA-Saclay. AM was supported in part by a TALENT fellowship awarded by CY Cergy Paris Universit\'e. SR was supported in part by a J C Bose Fellowship of the SERB, India.}


\clearpage
\title{Dynamics of Ordered Active Columns: Flows, Twists, and Waves\\ Supplementary Material}

\affiliation{Centre for Condensed Matter Theory, Department of Physics, Indian Institute of Science, Bangalore 560 012, India}

\affiliation{Department of Physics, Gibbet Hill Road, University of Warwick, Coventry CV4 7AL, United Kingdom}

\affiliation{{Laboratoire de Physique Th\'eorique et Mod\'elisation, CNRS UMR 8089,
		CY Cergy Paris Universit\'e, F-95032 Cergy-Pontoise Cedex, France}}
\affiliation{Sorbonne Universit\'{e} and CNRS, Laboratoire Jean Perrin, F-75005, Paris, France}

\affiliation{Centre for Condensed Matter Theory, Department of Physics, Indian Institute of Science, Bangalore 560 012, India}

\begin{abstract}
	In this supplement, we present detailed calculations corresponding to the results of the main text. We derive the equations of motion of a apolar, polar and chiral active columnar liquid crystals in Sec. \ref{colum_der}. We then show that effect of the lowest order achiral active stress in a columnar material is \emph{exactly} equivalent to that of an externally imposed stress in Sec. \ref{equiv} which allows us to control active columnar systems. Sec. \ref{SecHH} describes a spontaneous Helfrich-Hurault instability in active columnar materials.
	We describe the hydrodynamics of apolar and polar (but achiral) active columnar systems in Sec. \ref{act_ach}. We describe the hydrodynamics of chiral but apolar columnar structures in Sec. \ref{act_apol}. In Sec. \ref{act_pol}, we describe the novel odd-elastic and nonreciprocal response and wave propagation in polar and chiral columnar materials. Finally, in Sec. \ref{densfluc}, we show that density fluctuations in active columnar materials remain ``normal''.
\end{abstract}

\maketitle
\onecolumngrid
\appendix
\setcounter{equation}{0}
\section{Derivation of hydrodynamic equations of columnar materials starting from a theory in terms of phase fields}
\label{colum_der}
The dynamics of columnar fluids can be consistently derived from that of a conserved or non-conserved scalar field coupled to a velocity field. In an earlier paper, we discussed an analogous construction of the hydrodynamic equations for a layered system \cite{SJ_chiral_layered} starting with a scalar field theory and examining fluctuations about a state with a one-dimensional density modulation and now repeat the same construction for a columnar one. We will only ``derive'' the dynamical equations starting from a conserved scalar field coupled to a velocity field, but an exactly equivalent calculation can be performed for a non-conserved scalar field.

We consider a bulk, three-dimensional system with three-dimensional density $\psi({\bf r}, t)$ and velocity ${\bf v}({\bf r}, t)$ fields where ${\bf r}=(x,y,z)$. A columnar state is realised in this system when the density field has a two-dimensional, periodic density modulation. We will further examine the breaking of two discrete symmetries: chirality and polarity. The latter implies breaking the up-down symmetry in the fluid direction, i.e., the direction transverse to the density modulations. We will therefore examine four distinct kinds of active columnar liquid crystal, listed here from the most to the least symmetric: i. apolar and achiral; ii. polar and achiral; iii. apolar and chiral; and iv. polar and chiral. Note that the hydrodynamics of active, polar but achiral columnar liquid-crystalline phases is exactly equivalent to their apolar counterparts. Therefore, we will not need to consider i. and ii. separately.

Before embarking on deriving the hydrodynamic equations, it is useful to comment on the distinct ways in which polar asymmetry appears in columnar materials and lamellar phases. Here, we discuss systems that have polar asymmetry in the fluid direction, i.e., orthogonal to the periodic directions. Polar lamellar or smectic phases \cite{Chen_Toner} usually refer to systems in which the polar asymmetry is \emph{along} the normal to the layers i.e., along the direction of density modulation. A closer analogue to polar columnar phases are what are known as smectic-P phases \cite{Toner_smecticP}, which have a direction in the fluid plane of the layers. However, since in three-dimensional lamellar phases, there are two fluid directions, a polar asymmetry in the fluid directions requires breaking a continuous symmetry; two-dimensional smectic-P states do break only a discrete symmetry in the fluid direction like three-dimensional polar columnar liquid crystals we will discuss. This distinction between polar lamellar and columnar materials has certain interesting consequences. For instance, while the hydrodynamics of polar and apolar smectic phases in Stokesian fluids are equivalent, just like their columnar counterparts, the hydrodynamics of ``dry'' polar and active smectics \cite{ChenToner} -- i.e. compressible polar and active smectics moving through a frictional media (in arbitrary dimensions) -- is distinct from dry, active apolar lamellar phases \cite{Tap_smec}.
There is no such distinction between a dry polar and apolar columnar system. Since the polar drive in columnar systems is \emph{orthogonal} to the plane in which translation symmetry is broken, the displacement field dynamics is unaffected, to the lowest order in wavenumbers, by this drive.

With this preamble, we now construct the hydrodynamic equations for active columnar liquid crystals.

\subsection{Apolar and achiral active columnar phase}
\label{apolSec}
We start with the general dynamics of a conserved scalar field coupled to a velocity field -- an active version of model H \cite{CatesH1, CatesH2, HalpHohen}
\begin{equation}
	\label{psieq3d1}
	\partial_t\psi=-\nabla\cdot({\psi}{\bf v})+M\nabla^2\frac{\delta {F}[\psi]}{\delta\psi}+\nabla\cdot(\psi{\bf V}_{an})
	+{\xi}_\psi,
\end{equation}
where $\langle\xi_\psi({\bf r}, t)\xi_\psi({\bf r}', t')\rangle=-2D\nabla^2\delta({\bf r}-{\bf r}')\delta(t-t')$ 
and the active ``particle-phase'' velocity is 
\begin{equation}
	\label{PPvel}
	{\bf V}_{an}=\mu_1(\nabla\psi)\nabla^2\psi+\mu_2\nabla\cdot(\nabla\psi\nabla\psi)+\mu_3\nabla(\psi\nabla^2\psi).
\end{equation}
The velocity field, in the limit of small Reynolds number relevant for most biological and soft matter systems, has an overdamped, Stokesian dynamics:
\begin{equation}
	\label{vel3dpsi1}
	\eta\nabla^2{v}_i=\psi\partial_i\frac{\delta F[\psi]}{\delta \psi}+\partial_i p-{\zeta}_H\partial_j(\partial_i\psi\partial_j\psi)
	+\xi_{v_i},
\end{equation}
where $p$ is the pressure that enforces the incompressibility constraint $\nabla\cdot{\bf v}=0$ and $\langle{{\xi}}_{v_i}({\bf r},t){{\xi}}_{v_j}({\bf r}',t')\rangle=-2\delta_{ij}{D}_v{\nabla}^2\delta({\bf r}-{\bf r}')\delta(t-t')$. The force density with a coefficient $\zeta_H$ is one of the active terms in this model. The constraint of detailed balance would also introduce a relation between the coefficients $\mu_i$ in \eqref{PPvel}.
Further, in equilibrium, the noise strengths are related via ${D}/{M}={D}_v/{\eta}$.

We now separate the field $\psi$ into a part $\psi_0$ that is uniform on average and a part $\psi_1$ that is modulated on average and use a Swift-Hohenberg free energy to describe the latter.  
The Swift-Hohenberg free energy is given by 
\begin{equation}
	F=\frac{\Upsilon}{2}\int_{\bf x}\left[\frac{\alpha}{2}\psi_1^2+\frac{\beta}{4}\psi_1^4+\left |\{\nabla^2+q^2_{s}\}\psi_1\right |^2\right] ,
\end{equation}
where $q_s$ sets the lengthscale of the spatial modulation. 
To obtain the dynamics of a columnar material from this, we assume a density profile that is periodic in two-dimensions.
\begin{equation}
	{\psi}_1|_{s.s}=\sum_j\psi^0_j [e^{i\phi_j}+e^{-i\phi_j}],
\end{equation}
where
\begin{equation}
	\phi_j=q_s({\bf q}_j\cdot{\bf x}-\bar{{\bf q}}_j\cdot{\bf u}_\perp)
\end{equation}
with ${\bf q}_j$ being the lattice vectors of the planar structure and ${\bf u}_\perp$ is the two-dimensional displacement field about the periodically ordered density wave state. We can take the lattice vectors to be $(-\sqrt{3}/2,-1/2,0)$, $(\sqrt{3}/2,-1/2,0)$ and $(0,1,0)$ for a triangular structure without loss of generality. We write the corresponding two-dimensional vectors as $\bar{{\bf q}}_1=(-\sqrt{3}/2,-1/2)$, $\bar{{\bf q}}_2=(\sqrt{3}/2,-1/2)$ and $\bar{{\bf q}}_3=(0,1)$. 
Taking all $\psi^0_j$ to be equal to $\psi^0_1=\sqrt{|\alpha|/\beta}$, we obtain a free energy density in terms of the displacement field alone:
\begin{equation}
	\label{fenerg1}
	f_{{\bf u}_\perp}=\frac{1}{2}\left[\lambda\text{Tr}[{\bsf E}]^2+2\mu{\bsf E}:{\bsf E}+K\nabla^2u_k\nabla^2u_k \right]
\end{equation}
where we have defined a rotation-invariant strain tensor
\begin{multline}
	E_{ij}=\frac{1}{2}[\partial_iu_j+\partial_ju_i-\partial_iu_k\partial_ju_k-\partial_zu_i\partial_zu_j]\\=\frac{1}{2}\begin{pmatrix}
		2\partial_x u_x-(\partial_x u_x)^2-(\partial_x u_y)^2-(\partial_z u_x)^2 & \partial_x u_y+\partial_y u_x-\partial_x u_x\partial_y u_x-\partial_xu_y\partial_y u_y-\partial_zu_x\partial_zu_y\\\partial_x u_y+\partial_y u_x-\partial_x u_x\partial_y u_x-\partial_xu_y\partial_y u_y-\partial_zu_x\partial_zu_y &2\partial_y u_y-(\partial_y u_x)^2-(\partial_y u_y)^2-(\partial_z u_y)^2
	\end{pmatrix},
\end{multline}
and $\lambda=3\Upsilon{\psi^0_1}^2q_s^4$, $\mu=3\Upsilon{\psi^0_1}^2q_s^4$ and $K=3\Upsilon{\psi^0_1}^2q_s^2$. Note that, as pointed out by Kamien in the context of smectics \cite{Kamien}, the \emph{nonlinear} strain tensor is \emph{not unique}. For instance, we could have equivalently defined another strain tensor
\begin{equation}
	\bar{E}_{ij}=\frac{1}{2}[\partial_iu_j+\partial_ju_i-\partial_ku_i\partial_ku_j]
\end{equation}
in terms of which, the free energy would have the form
\begin{equation}
	\bar{f}_{{\bf u}_\perp}=\frac{1}{2}\left[\bar{\lambda}\text{Tr}[\bar{{\bsf E}}]^2+\bar{\mu}\tilde{\bar{{\bsf E}}}:\tilde{\bar{{\bsf E}}}+\bar{K}\nabla^2u_k\nabla^2u_k \right]
\end{equation}
where $\tilde{\bar{{\bsf E}}}$ is the deviatoric part of the strain tensor, $\bar{\lambda}=6\Upsilon{\psi^0_1}^2q_s^4$, $\bar{\mu}=6\Upsilon{\psi^0_1}^2q_s^4$ and $\bar{K}=6\Upsilon{\psi^0_1}^2q_s^2$. 

With the definition of free energy density $f_{{\bf u}_\perp}$ and the strain tensor $E_{ij}$ in hand, we are ready to construct the equations of motion of the columnar phase in terms of the displacement field about a periodic, two-dimensional density profile. For this, we will use the following relations:
\begin{equation}
	\nabla\psi_1\nabla\psi_1=2{\psi^0_1}^2\sum_j\nabla\phi_j\nabla\phi_j
\end{equation}
Then, using the definition of $\phi_j$,
\begin{equation}
	\sum_j\nabla_l\phi_j\nabla_k\phi_j=q_s^2\sum_jq_{j_i}q_{j_m}[\delta_{ki}-\nabla_ku_i][\delta_{ml}-\nabla_lu_m]=\frac{3q_s^2}{2}[\delta_{ki}-\nabla_ku_i][\delta_{li}-\nabla_lu_i]
\end{equation}
Using the dynamical equations for $\psi$, we get a dynamical equation for ${\bf u}_\perp$:
\begin{equation}
	\partial_t{\bf u}_\perp={\bf v}_\perp-{\bf v}\cdot\nabla{\bf u}_\perp-\Gamma\frac{\delta F_{{\bf u}}}{\delta{\bf u}_\perp}+{{\bf V}_{a}}_\perp,
\end{equation}
where $\Gamma\propto Mq_s^2$ and ${{\bf V}_{a}}_\perp$ is the active permeation which, for simplicity, we only write to linear order:
\begin{equation}
	{{\bf V}_{a}}_\perp=\mu_1\nabla^2{\bf u}_\perp+\mu_2\nabla_\perp(\nabla_\perp\cdot{\bf u}_\perp),
\end{equation}
with $\mu_1=-3q_s^2\psi_0^2(\mu_2+\mu_1)$ and $\mu_2=-3q_s^2\psi_0^2(\mu_2-2\mu_3)$.

The velocity of the mass-density wave -- described by the dynamics of the displacement field -- is not the same as the joint velocity of the fluid and the mass-density wave. This difference is captured by the active and passive permeation term ${\delta F_{{\bf u}}}/{\delta{\bf u}_\perp}$ and ${\bf V}_{a\perp}$ in the ${\bf u}_\perp$ equation. However, for fluctuations along most directions of the wavevector space (but not all!), the effect of these permeative terms will turn out to be subdominant. 

Now, we construct the dynamical equation for the Stokesian velocity field. This has the form 
\begin{equation}
	\eta\nabla^2{v}_i=\partial_i\Pi+(\delta_{ij}-\partial_i{u_\perp}_j)\frac{\delta F_{{\bf u}}}{\delta {u_\perp}_j}
	+\nabla\cdot\bm{\sigma}^a,
\end{equation}
where $\bm{\sigma}^a$ is the active stress and $\Pi$ is the pressure enforcing the incompressibility constraint. The full, nonlinear active stress is 
\begin{equation}
	\label{nlinstr}
	\zeta_H\partial_i\psi\partial_j\psi=3\zeta_H{\psi_1^0}^2q_s^2\begin{pmatrix}(\partial_x u_x-1)^2+(\partial_x u_y)^2 & \partial_yu_x(\partial_ xu_x-1)+(\partial_yu_y-1)\partial_xu_y & \partial_zu_x(\partial_x u_x-1)+\partial_zu_y\partial_xu_y\\\partial_yu_x(\partial_xu_x-1)+\partial_xu_y(\partial_yu_y-1) &(\partial_y u_x)^2+(\partial_yu_y-1)^2 &\partial_zu_x\partial_yu_x+\partial_zu_y(\partial_yu_y-1)\\\partial_zu_x(\partial_xu_x-1)+\partial_zu_y\partial_xu_y&\partial_zu_x\partial_yu_x+\partial_zu_y(\partial_yu_y-1) &(\partial_z u_x)^2+(\partial_zu_y)^2\end{pmatrix}
\end{equation}
Note that this stress is invariant under arbitrary rotations, by definition. 
The linearised force balance equation can, therefore, be explicitly written as
\begin{equation}
	\eta\nabla^2{\bf v}=\nabla\Pi-(\mu+\zeta)\nabla_\perp^2{\bf u}_\perp-(\lambda+\mu+\zeta)\nabla_\perp\nabla_\perp\cdot{\bf u}_\perp-\zeta\partial_z^2{\bf u}_\perp-\zeta\partial_z(\nabla_\perp\cdot{\bf u}_\perp)\hat{{\bf z}},
\end{equation}
where $\zeta=-3\zeta_H{\psi_1^0}^2q_s^2$.
Further, noting that the joint velocity field is incompressible and accordingly redefining the pressure, which is merely a Lagrange multiplier to enforce the incompressibility constraint, the active force is seen to be simply $\zeta\nabla^2{\bf u}_\perp$ i.e.,
\begin{equation}
	\eta\nabla^2{\bf v}=\nabla\Pi-\mu\nabla_\perp^2{\bf u}_\perp-(\lambda+\mu)\nabla_\perp\nabla_\perp\cdot{\bf u}_\perp-\zeta\nabla^2{\bf u}_\perp.
\end{equation}

However, this is not the most general hydrodynamic theory of an active columnar system.
In particular, the coefficient of the active stress can be anisotropic since the columnar state itself is anisotropic. The easiest way to account for this anisotropy is to retain an explicit director field $\hat{n}$ and start with the columnar and active variant of the Chen-Lubensky model \cite{ChenLub}. For this, we replace the Swift-Hohenberg free energy with $F=\int_{\bf r}[f_{\psi_0}+f_n+f_{\psi_1, n}]$, where $f_n$ is the free energy density for the director field, and
\begin{equation}
	\label{psinfenerg}
	f_{\psi_1, n}=\frac{\Upsilon}{2}\left[\frac{\alpha}{2}\psi_1^2+\frac{\beta}{4}\psi_1^4+C_\parallel (\hat{n}\cdot\nabla\psi_1)^2+\left |\{(\nabla-\hat{n}\hat{n}\cdot\nabla)^2+q^2_{s}\}\psi_1\right |^2\right].
\end{equation}
Taking the mean director direction to be $\hat{n}=\hat{{\bf z}}$ and using the same $\psi_1$ as earlier, we obtain a free energy density containing both ${\bf u}_\perp$ and $\hat{n}$
\begin{equation}
	f_{{\bf u}_\perp,\delta{\bf n}_\perp}=\frac{1}{2}\left[\lambda\text{Tr}[{\bsf E}]^2+2\mu{\bsf E}:{\bsf E}+K\nabla^2u_k\nabla^2u_k +C(\delta{n}_\perp-\partial_z{\bf u}_\perp)^2\right]
\end{equation}
where $C=3\Upsilon C_\parallel{\psi^0_1}^2q_s^2$. This free energy implies that the \emph{fluctuations} of the director are slaved to that of the displacement field. 
That is, the director field is not a hydrodynamic quantity and relaxes fast to 
\begin{equation}
	\label{dir}
	\hat{n}=\frac{(\partial_z{\bf u}_\perp,1)}{\sqrt{1+\partial_z{\bf u}_\perp\cdot\partial_z{\bf u}_\perp}}\approx(\partial_z{\bf u}_\perp,1)
\end{equation}
where the approximate equality ignores higher order nonlinear contributions. We can integrate out the director fluctuations from the free energy to obtain an expression purely in terms of the displacement which has the form \eqref{fenerg1}.

In general, the active stress also depends on $\hat{n}$; i.e., it is anisotropic. This can be seen by, for instance, writing the stress as
\begin{equation}
	\label{anisostr}
	\sigma^a_{ij}=\zeta_{ijkl}\partial_k\psi\partial_l\psi
\end{equation}
where
\begin{equation}
	\label{aniso}
	\zeta_{ijkl}=\zeta_1n_in_jn_kn_l+\zeta_2(\delta_{ik}\delta_{jl}+\delta_{il}\delta_{jk})+\zeta_3(n_in_k\delta_{jl}+n_jn_k\delta_{il}+n_in_l\delta_{jk}+n_jn_l\delta_{ik})+\zeta_4\delta_{ij}\delta_{kl}+\zeta_5(\delta_{ij}n_kn_l+n_in_j\delta_{kl}).
\end{equation}

To calculate the linearised active force density, we first obtain the $\sigma_{ij}$ using \eqref{anisostr}, \eqref{aniso}, \eqref{dir} and \eqref{nlinstr} and expand to first order in the displacement fields.
The divergence of this yields the achiral active force density as
\begin{equation}
	{\bf f}^a=\zeta_2\nabla^2{\bf u}_\perp-2\zeta_5\partial_z^2{\bf u}_\perp,
\end{equation}
where we have dropped a term $\propto \nabla\nabla_\perp\cdot{\bf u}_\perp$ since it can be absorbed in a redefinition of the pressure.
Defining $\zeta_\perp=\zeta_2$ and $\zeta_\parallel=(\zeta_2-2\zeta_5)$, the linearised force balance equation, after eliminating terms that can be absorbed into a redefinition of the pressure, becomes
\begin{equation}
	\eta\nabla^2{\bf v}=\nabla\Pi-(\mu+\zeta_\perp)\nabla_\perp^2{\bf u}_\perp-(\lambda+\mu)\nabla_\perp\nabla_\perp\cdot{\bf u}_\perp-\zeta_\parallel\partial_z^2{\bf u}_\perp.
\end{equation}
The careful reader may notice that we could have introduced another active stress $\propto n_in_j$ which would lead to a force density $\partial_z^2{\bf u}_\perp+\hat{{\bf z}}\partial_z\nabla_\perp\cdot{\bf u}_\perp$. Using incompressibility, to redefine the pressure, we see that this force density renormalises the coefficient of $\nabla_\perp\nabla_\perp\cdot{\bf u}_\perp$ and $\partial_z^2{\bf u}_\perp$; i.e., it also leads to an active modification of the bulk modulus. In general, in an incompressible columnar system, the \emph{only} active force-density that cannot be reabsorbed by an activity-dependent renormalisation of the shear and the bulk moduli, at this order in gradients, is $\partial_z^2{\bf u}_\perp$. In compressible materials, one would have a further purely active contribution to the force density $\propto\partial_z\nabla_\perp\cdot{\bf u}_\perp$. 

Similarly, the active permeation terms can also be anisotropic. However, as we shall show, they become important only along one specific direction. Therefore, we do not examine their anisotropy here.

\subsection{Apolar and chiral active columnar phase}
The dynamical equations for $\psi$ and the velocity field for a \emph{chiral} columnar phase have the following additional \emph{chiral} terms: i. Eq. \eqref{psieq3d1} is supplemented by the divergence of a \emph{passive} chiral current $\nabla\cdot{\bf J}_c$ and ii. Eq. \eqref{vel3dpsi1} is supplement both by a passive chiral stress which is the Onsager counterpart to ${\bf J}_c$ \cite{SJ_chiral_layered, Andreev} and an active chiral stress $\bm{\sigma}_c$. The chiral current is
\begin{equation}
	{\bf J}_c=\Omega_v\psi\nabla^2(\nabla\times{\bf v}).
\end{equation}
This term distinguishes the {dynamics} of \emph{passive} achiral and chiral columnar phases, which are indistinguishable at the hydrodynamic level, at higher order in wavenumbers. However, this term yields velocity couplings $\propto\nabla^2(\partial_z v_y, -\partial_z v_x)$ in the displacement field dynamics which do not affect the long-time, large-scale dynamics. Therefore, we will ignore both this term and its Onsager (anti)symmetric counterpart in the force balance equation. The chiral active stress in \eqref{vel3dpsi1} has the form
\begin{equation}
	\sigma^{ac}_{ij}=\bar{z}_c \partial_l(\epsilon_{ijk}\partial_k\psi\partial_l\psi)
\end{equation}
where $\epsilon_{ijk}$ is the \emph{three-dimensional} Levi-Civita tensor. The chiral active force-density, with a coefficient $\bar{z}_c $, appears at a \emph{higher} order in gradients than the achiral active force. This force has a form similar to the one used in theories of three-dimensional chiral nematic liquid crystals \cite{Seb_1, furthauer2012, Cates_drop, Cates_Marko} and the velocity field resulting from it can be shown to be divergence-free. 
Note that though this stress superficially seems to be antisymmetric, it is allowed in angular momentum-conserved systems; in fact, an equivalent (up to a Belinfante-Rosenfeld tensor) explicitly \emph{symmetric} stress yielding the same velocity field can be constructed \cite{MPP, Cates_Marko}:
\begin{equation}
	\bar{z}_c \partial_l(\epsilon_{ijk}\partial_k\psi\partial_l\psi)\equiv \bar{z}_c [\epsilon_{ilk}\partial_l(\partial_k\psi\partial_j\psi)+\epsilon_{jlk}\partial_l(\partial_k\psi\partial_i\psi)].
\end{equation}

In terms of the displacement field, the linearised chiral active stress becomes
\begin{equation}
	\bar{z}_c\epsilon_{ijk}\partial_l(\partial_l\psi\partial_k\psi)=-3\bar{z}_c\psi_0^2q_s^2\begin{pmatrix}  0& \partial_z\nabla_\perp\cdot{\bf u}_\perp &-(\partial_z^2 u_y+\nabla_\perp^2u_y+\partial_y\nabla_\perp\cdot{\bf u}_\perp)\\-\partial_z\nabla_\perp\cdot{\bf u}_\perp & 0& \partial_z^2u_x+\nabla_\perp^2u_x+\partial_x\nabla_\perp\cdot{\bf u}_\perp\\(\partial_z^2 u_y+\nabla_\perp^2u_y+\partial_y\nabla_\perp\cdot{\bf u}_\perp) &-(\partial_z^2u_x+\nabla_\perp^2u_x+\partial_x\nabla_\perp\cdot{\bf u}_\perp) & 0\end{pmatrix}.
\end{equation}
Defining $\zeta_c=-3\bar{z}_c{\psi^0_1}^2q_s^2$, this implies that the force balance equation for a chiral, apolar columnar material is 
\begin{equation}
	\eta\nabla^2{\bf v}=\nabla\Pi-(\mu+\zeta)\nabla_\perp^2{\bf u}_\perp-(\mu+\lambda+\zeta)\nabla_\perp\nabla_\perp\cdot{\bf u}_\perp-\zeta\partial_z^2{\bf u}_\perp-\zeta\partial_z(\nabla_\perp\cdot{\bf u}_\perp)-\zeta_c\nabla^2\partial_z\boldsymbol{\epsilon}\cdot{\bf u}_\perp+\zeta_c\nabla^2\nabla_\perp\times{\bf u}_\perp,
\end{equation}
where $\boldsymbol{\epsilon}$ is the \emph{rank-two} Levi-Civita tensor. More compactly, 
\begin{equation}
	\eta\nabla^2{\bf v}=\nabla\Pi-\mu\nabla_\perp^2{\bf u}_\perp-(\mu+\lambda)\nabla_\perp\nabla_\perp\cdot{\bf u}_\perp-\zeta\nabla^2{\bf u}_\perp+\zeta_c\nabla^2\nabla\times{\bf u}_\perp.
\end{equation}

Of course, as in the last section, the chiral active stress may also be anisotropic. One (non-unique) version of this is obtained by writing
\begin{equation}
	\sigma^{ac}_{ij}=({{\bar{z}_c}})_{ijkl} [\epsilon_{kmn}\partial_m(\partial_n\psi\partial_l\psi)+\epsilon_{lmn}\partial_m(\partial_n\psi\partial_k\psi)],
\end{equation}
where we have used an explicitly symmetrised version of the chiral stress and
\begin{equation}
	-3{\psi^0_1}^2q_s^2({{\bar{z}_c}})_{ijkl}={\zeta_c}_1n_in_jn_kn_l+{\zeta_c}_2(\delta_{ik}\delta_{jl}+\delta_{il}\delta_{jk})+{\zeta_c}_3(n_in_k\delta_{jl}+n_jn_k\delta_{il}+n_in_l\delta_{jk}+n_jn_l\delta_{ik})+{\zeta_c}_4\delta_{ij}\delta_{kl}+{\zeta_c}_5(\delta_{ij}n_kn_l+n_in_j\delta_{kl}).
\end{equation}
Again taking $\hat{n}\approx(\partial_z{\bf u}_\perp,1)$, we get the  anisotropic force balance equation for a chiral columnar fluid:
\begin{multline}
	\eta\nabla^2{\bf v}=\nabla\Pi-(\mu+\zeta_\perp)\nabla_\perp^2{\bf u}_\perp-(\lambda+\mu)\nabla_\perp\nabla_\perp\cdot{\bf u}_\perp-\zeta_\parallel\partial_z^2{\bf u}_\perp-[\zeta^c_1\partial_z^2+\zeta^c_2\nabla_\perp^2][\partial_z(\boldsymbol{\epsilon}\cdot{\bf u}_\perp)]+[\zeta^c_3\partial_z^2+\zeta^c_4\nabla_\perp^2)](\nabla_\perp\times{\bf u}_\perp),
\end{multline}
where we have defined $\zeta^c_1=(2\zeta_{c2}+\zeta_{c3})$, $\zeta^c_2=2\zeta_{c2}$, $\zeta^c_3=(4\zeta_{c1}+2\zeta_{c2}+2\zeta_{c3})$ and $\zeta^c_4=(2\zeta_{c2}+\zeta_{c3})$.

\subsection{Polar and chiral active columnar phase}
\label{polcolsec}
In this section, we will construct the hydrodynamic equations for an active, polar and chiral columnar material. 
It is tempting to obtain the equations of motion of a chiral and polar columnar phase starting from a variant of model $H$ which lacks $\psi\to -\psi$ symmetry. However, this is not possible since the polar asymmetry is along the fluid direction.
We will obtain the dynamics by starting with a model analogous to the one which we used to obtain the anisotropic apolar columnar dynamics in Sec. \ref{apolSec} but with a polar vector ${\bf P}$ instead of a director.
The free energy is $F=\int_{\bf r}[f_{\psi_0}+f_P+f_{\psi_1, P}]$, where $F_P$ is the Landau-de Gennes free energy density for the polar order parameter, and
\begin{equation}
	f_{\psi_1, P}=\frac{\Upsilon}{2}\left[\frac{\alpha}{2}\psi_1^2+\frac{\beta}{4}\psi_1^4+C_\parallel (\hat{P}\cdot\nabla\psi_1)^2+\left |\{(\nabla-\hat{P}\hat{P}\cdot\nabla)^2+q^2_{s}\}\psi_1\right |^2\right].
\end{equation}
The density wave state is assumed to arise in this case in a phase with a mean ${\bf P}$. Taking the mean ${\bf P}=P_0\hat{{\bf z}}$ and assuming a two-dimensional density wave state as in Sec.  \ref{apolSec},
we get from $f_{\psi_1, P}$
\begin{equation}
	f_{{\bf u}_\perp,\delta{\bf P}_\perp}=\frac{1}{2}\left[\lambda\text{Tr}[{\bsf E}]^2+2\mu{\bsf E}:{\bsf E}+K\nabla^2u_k\nabla^2u_k +C(\delta{\bf P}_\perp-\partial_z{\bf u}_\perp)^2\right]
\end{equation}
where $C=3\Upsilon C_\parallel{\psi^0_1}^2q_s^2$. This free energy implies that the {fluctuations} of the polarisation vector are slaved to that of the displacement field. Therefore, in the following, we construct the dynamical equations of active and chiral polar columnar phases in terms of the displacement field ${\bf u}_\perp$ and the velocity field ${\bf v}$ just as in the earlier sections. The polarisation vector is assumed to relax fast to ${\bf P}=P_0(\partial_z{\bf u}_\perp,1)$. Note that the free energy of polar columnar liquid crystals is exactly equivalent to their apolar counterparts. 

We now construct an active stress that is both \emph{polar} and \emph{chiral},
\begin{equation}
	\label{polchistrs}
	\sigma^{pc}_{ij}=\bar{\zeta}_{pc}[\epsilon_{ikl}P_l\partial_k\psi\partial_j\psi+\epsilon_{jkl}P_l\partial_k\psi\partial_i\psi],
\end{equation}
which is symmetric by construction. Defining $\zeta_{pc}=-3q_s^2{\psi^0_1}^2P_0\bar{\zeta}_{pc}$, we obtain the simplest chiral and polar active stress contribution in columnar materials in terms of the displacement fields from this;
the linear part of this stress is 
\begin{equation}
	\left[\bm{\sigma}^{pc}\right]_{lin}=\zeta_{pc}\begin{pmatrix}\partial_x u_y+\partial_y u_x& \partial_y u_y-\partial_x u_x &0\\\partial_y u_y-\partial_x u_x& -(\partial_x u_y+\partial_y u_x) &0\\ 0 &0 &0\end{pmatrix}.
\end{equation}
This \emph{linearised} stress is \emph{exactly} equivalent to the odd elastic stress of momentum-conserved two-dimensional chiral solids \cite{Ano_hex, Deboo2}. 
The difference between the polar, chiral stress in columnar materials and the odd elastic stress in two-dimensional chiral solids appears only at the nonlinear level; these are irrelevant for large-scale, long-time dynamics of the system and, therefore, we do not explicitly display them. The stress $\bm{\sigma}^{pc}$ has this form because of a combination of rotation invariance and momentum conservation. It has to be constructed out of the strain tensor, the three-dimensional Levi-Civita tensor and the polarisation. Since the strain tensor in a columnar system is two-dimensional, the only (linearised) odd elastic stress that one can construct is equivalent to that in \cite{Ano_hex, Deboo2}.

The general force balance equation for a polar and chiral active columnar system is
\begin{equation}
	\eta\nabla^2{\bf v}=\nabla\Pi-(\mu+\zeta_\perp)\nabla_\perp^2{\bf u}_\perp-(\lambda+\mu)\nabla_\perp\nabla_\perp\cdot{\bf u}_\perp-\zeta_\parallel\partial_z^2{\bf u}_\perp-\zeta_{pc}\nabla_\perp^2\boldsymbol{\epsilon}\cdot{\bf u}_\perp,
\end{equation}
where we have discarded higher order in gradient contributions.

\section{Active columnars: Effect of an isotropic stress} 
\label{equiv}
In this section, we demonstrate that the effect of the active achiral and apolar stress in an incompressible columnar system can be \emph{exactly} compensated by an external stress. To show this, we note that
the nonlinear strain tensor in a columnar is 
\begin{equation}
	{\bsf E}=\frac{1}{2}\begin{pmatrix}
		2\partial_x u_x-(\partial_x u_x)^2-(\partial_x u_y)^2-(\partial_z u_x)^2 & \partial_x u_y+\partial_y u_x-\partial_x u_x\partial_y u_x-\partial_xu_y\partial_y u_y-\partial_zu_x\partial_zu_y\\\partial_x u_y+\partial_y u_x-\partial_x u_x\partial_y u_x-\partial_xu_y\partial_y u_y-\partial_zu_x\partial_zu_y &2\partial_y u_y-(\partial_y u_x)^2-(\partial_y u_y)^2-(\partial_z u_y)^2
	\end{pmatrix}
\end{equation}
and the free energy without an external stress is 
\begin{equation}
	F=\frac{1}{2}\int [\lambda E_{ii}^2+2\mu E_{ij}E_{ij}].
\end{equation}
A purely compressive stress leads to the extra free energy term 
\begin{equation}
	F_{\text{ext}}=\int \sigma_0E_{ii}=\int \sigma_0E
\end{equation}
where we have denoted $\text{Tr}[{\bsf E}]$ by 
\begin{equation}
	E=\nabla_l u_l-\frac{1}{2}\nabla_m u_l\nabla_mu_l.
\end{equation}
(Note that exactly the same calculation (with the same result) can also be performed in terms of the strain tensor $\bar{\bsf{E}}$.) 
The force due to this external stress is 
\begin{equation}
	{\bf f}_{ext}=-(\delta_{ij}-\nabla_iu_j)\frac{\delta F_{\text{ext}}}{\delta u_j}=\sigma_0(\delta_{ij}-\nabla_iu_j)\nabla_k\frac{\partial E}{\partial \nabla_k u_j}
\end{equation}
Using the definition of $E$, 
\begin{equation}
	\frac{\partial E}{\partial \nabla_k u_j}=\delta_{kj}-\nabla_ku_j,
\end{equation}
which implies 
\begin{multline}
	(\delta_{ij}-\nabla_iu_j)\nabla_k\frac{\partial E}{\partial \nabla_k u_j}=\nabla_k[	(\delta_{ij}-\nabla_iu_j)(\delta_{kj}-\nabla_ku_j)]-(\delta_{kj}-\nabla_ku_j)\nabla_k(\delta_{ij}-\nabla_iu_j)\\=\nabla_k\left[\frac{\partial E}{\partial \nabla_i u_j}\frac{\partial E}{\partial \nabla_k u_j}\right]-(\delta_{kj}-\nabla_ku_j)\nabla_k(\delta_{ij}-\nabla_iu_j)
\end{multline}
\begin{equation}
	(\delta_{kj}-\nabla_ku_j)\nabla_k(\delta_{ij}-\nabla_iu_j)=-[\nabla_i\nabla_j u_j-(\nabla_ku_j)\nabla_i(\nabla_k u_j)]=-\nabla_i\left[\nabla_ju_j-\nabla_k u_j\nabla_ku_j\right]=-\nabla E
\end{equation}
This implies that 
\begin{equation}
	{\bf f}_{ext}=\sigma_0\nabla_k\left[\frac{\partial E}{\partial \nabla_i u_j}\frac{\partial E}{\partial \nabla_k u_j}+E\delta_{ik}\right]
\end{equation}

The active, achiral stress is 
\begin{equation}
	\boldsymbol{\sigma}^a\propto\partial_k\psi\partial_l\psi=3q_s^2{\psi^0_1}^2[\delta_{ki}-\nabla_ku_i][\delta_{li}-\nabla_lu_i].
\end{equation}
This implies that
\begin{equation}
	\boldsymbol{\sigma}^a=\zeta_H\partial_k\psi\partial_l\psi=3\zeta_H q_s^2{\psi^0_1}^2\left[\frac{\partial E}{\partial \nabla_k u_m}\frac{\partial E}{\partial \nabla_l u_m}\right].
\end{equation}
That is, apart from a purely isotropic part which can be absorbed in a redefinition of the pressure, the active stress in a columnar system can be fully compensated by imposing an external stress of strength $\sigma_0=-3\zeta_H q_s^2{\psi^0_1}^2=\zeta$.

\section{Active Helfrich-Hurault instability }
\label{SecHH}
In this section, we use the mapping discussed in the last section between an active stress and an externally-imposed stress, to examine the active Helfrich-Hurault instability of apolar columnar liquid crystals. For this, we start with the well-known \emph{passive} Helfrich-Hurault under an external stress.
The rotation-invariant free energy for a columnar liquid crystal is 
\begin{equation}
	F = \int \frac{\lambda}{2} \biggl( \nabla\cdot{\bf u} - \frac{1}{2} \bigl| \nabla {\bf u} \bigr|^2 \biggr)^2 + \frac{K}{2} \bigl| \nabla^2 {\bf u} \bigr|^2 \,dV.
\end{equation}

As shown in the previous section, an active stress is equivalent to imposing an isotropic dilatational stress in a passive columnar phase. Such a dilatational stress corresponds leads to an Eulerian displacement field ${\bf u} = [\alpha x , \alpha y]$ where $\alpha$ is some positive constant. For this displacement field, the isotropic part of the strain field is
\begin{equation}
	\nabla\cdot{\bf u} - \frac{1}{2} \bigl| \nabla {\bf u} \bigr|^2 = 2 \alpha - \alpha^2 ,
\end{equation}
and the curvature is zero $\nabla^2 {\bf u} = 0$.  We consider a helical modulation of the columns about this strained state with the displacement field 
\begin{equation}
	{\bf u} = \alpha x \,{\bf e}_x + \alpha y \,{\bf e}_y + u_{0}(x,y) \bigl[ \cos qz \,{\bf e}_x + \sin qz \,{\bf e}_y \bigr].
\end{equation}
The strain and curvature are 
\begin{gather}
	\nabla\cdot{\bf u} - \frac{1}{2} \bigl| \nabla {\bf u} \bigr|^2 = 2 \alpha - \alpha^2 + (1-\alpha) \bigl[ \cos qz \,\partial_{x} u_0 + \sin qz \,\partial_{y} u_0 \bigr] - \frac{1}{2} \bigl| \nabla u_0 \bigr|^2 - \frac{1}{2} q^2 u_0^2 , \\
	\nabla^2 {\bf u} = \Bigl( \nabla^2 u_0 - q^2 u_0 \Bigr) \bigl[ \cos qz \,{\bf e}_x + \sin qz \,{\bf e}_y \bigr] .
\end{gather}
Substituting into the free energy and gathering terms quadratic in $u_0$ we find 
\begin{equation}
	F_{2} = \frac{\lambda}{2} \int \biggl\{ - \bigl( 2 \alpha - \alpha^2 \bigr) \Bigl[ \bigl| \nabla u_0 \bigr|^2 + q^2 u_0^2 \Bigr] + \frac{1}{2} (1-\alpha)^2 \bigl| \nabla u_0 \bigr|^2 + \gamma^2 \Bigl( \nabla^2 u_0 - q^2 u_0 \Bigr)^2 \biggr\} dV ,
	\label{eq:F2}
\end{equation}
where $\gamma = \sqrt{K/\lambda}$ is the `penetration depth'. 

We assume $u_0(x,y) = u_0 \cos(\pi x/L) \cos(\pi y/L)$ in the domain $[-L/2,L/2]^2\times H$ with a square cross-section and Dirichlet boundary conditions in the plane. With this choice, the quadratic free energy is 
\begin{equation}
	\begin{split}
		F_{2} & = \frac{\lambda L^2H u_0^2}{8} \biggl\{ - \bigl( 2 \alpha - \alpha^2 \bigr) \Bigl( 2 \bigl( \pi^2 / L^2 \bigr) + q^2 \Bigr) + (1-\alpha)^2 \bigl( \pi^2/L^2 \bigr) + \gamma^2 \Bigl( 2 \bigl( \pi^2 / L^2 \bigr) + q^2 \Bigr)^2 \biggr\} , \\
		& = \frac{\lambda L^2H u_0^2}{8} \biggl\{ \gamma^2 \biggl( 2 \bigl( \pi^2 / L^2 \bigr) + q^2 - \frac{\alpha - \frac{1}{2} \alpha^2}{\gamma^2} \biggr)^2  - \biggl( \frac{\alpha - \frac{1}{2} \alpha^2}{\gamma} \biggr)^2 + (1-\alpha)^2 \bigl( \pi^2/L^2 \bigr) \biggr\} ,
	\end{split}
\end{equation}
The instability threshold is obtained by examining when this quadratic energy first vanishes; this yields both the threshold strain for the \emph{passive} Helfrich-Hurault instability
and the selected length scale of the helical modulation just beyond the instability:
\begin{align}
	\alpha_{\mathrm{th}} & = 1 + \frac{\gamma \pi}{L} - \biggl[ 1 + \frac{\gamma^2\pi^2}{L^2} \biggr]^{1/2} \approx \frac{\gamma \pi}{L} , \\
	q_{\mathrm{th}}^2 & = \frac{\alpha - \frac{1}{2} \alpha^2}{\gamma^2} - \frac{2\pi^2}{L^2} \approx \frac{\pi}{\gamma L}.
\end{align}
We now use the stress-strain relation to convert the threshold strain $\alpha_{\mathrm{th}}$ to a threshold stress which must be applied on a passive columnar liquid crystal for it suffer a Helfrich-Hurault instability:
\begin{equation}
	\sigma_{0_{th}}=\dfrac{\sqrt{K\lambda}\pi}{L}.
\end{equation}
Given the mapping discussed in the last section, this is also the threshold \emph{active} stress $\zeta_{th}=-{\sqrt{K\lambda}\pi}/{L}$ at which an \emph{active} columnar liquid crystal undergoes a spontaneous Helfrich-Hurault instability.

\section{Linear dynamics of columnar systems}
In this section, we will discuss the linearised dynamics of active columnar systems. To construct the linear equations, we decompose the displacement fields into in-plane longitudinal and transverse displacement fields $u_l=(q_xu_x+q_yu_y)/\sqrt{q_x^2+q_y^2}$ and $u_t=(q_xu_y-q_yu_x)/\sqrt{q_x^2+q_y^2}$. Note that in a \emph{pure} two-dimensional poroelastic solid, two-dimensional incompressibility would imply that $\partial_tu_l\sim \mathcal{O}(q^2)$. However, since in this columnar system, the incompressibility constraint is three-dimensional while the displacement field is two-dimensional, $u_l$ has a $\mathcal{O}(q^0)$ relaxation rate except for \emph{purely} in-plane perturbations. We also write the wavevectors as $q_x=q\cos\phi\sin\theta$, $q_y=q\sin\phi\sin\theta$ and $q_z=q\cos\theta$. The white noises in the displacement dynamics appear from two sources i. from the velocity coupling and ii. the noise directly in the displacement field equation. Assuming the noises are such that they would lead to an equilibrium dynamics in the absence of active forces, the noise correlators for the longitudinal and transverse displacement fields are, respectively,
\begin{equation}
	\langle\xi_l({\bf q},\omega)\xi_l({\bf q}',\omega')\rangle=2T\left[\frac{\cos^2\theta}{\eta q^2}+\Gamma\right]\delta({\bf q}+{\bf q}')\delta(\omega+\omega'),
\end{equation}
\begin{equation}
	\langle\xi_t({\bf q},\omega)\xi_t({\bf q}',\omega')\rangle=2T\left[\frac{1}{\eta q^2}+\Gamma\right]\delta({\bf q}+{\bf q}')\delta(\omega+\omega')
\end{equation}
and
\begin{equation}
	\langle\xi_l({\bf q},\omega)\xi_t({\bf q}',\omega')\rangle=0.
\end{equation}
\subsection{Achiral and apolar active columnars}
\label{act_ach}
The linearised dynamics of the displacement fields, in this case, turn out to be
\begin{equation}
	\label{longapol}
	\partial_t u_l=-\cos^2\theta\left[\frac{2\mu+\lambda}{\eta}\sin^2\theta+\frac{\zeta_\perp}{\eta}\sin^2\theta+\frac{\zeta_\parallel}{\eta}\cos^2\theta\right]u_l+\xi_l
\end{equation}
and
\begin{equation}
	\partial_t u_t=-\left[\frac{\mu}{\eta}\sin^2\theta+\frac{\zeta_\perp}{\eta}\sin^2\theta+\frac{\zeta_\parallel}{\eta}\cos^2\theta\right]u_t+\xi_t.
\end{equation}
The longitudinal and the transverse displacement fluctuations are decoupled. Note that, as advertised, the $\mathcal{O}(q^0)$ relaxation rate in \eqref{longapol} vanishes for a fluctuation with wavevector with $\theta=0$, i.e., for a fluctuation purely in the crystalline plane. The relaxation rate of longitudinal displacement fluctuations for fluctuations with wavectors with $\theta=\pi/2$ will therefore be $\mathcal{O}(q^2)$ and be given by the permeative terms. From Sec. \ref{apolSec}, for a wavevector with $\theta=\pi/2$, the relaxation rate of $u_l$ is, therefore, $-[(2\mu+\lambda)\Gamma+2\mu_1+\mu_2]q^2$. For stability, therefore, we also require $[(2\mu+\lambda)\Gamma+2\mu_1+\mu_2]>0$ in addition to $\zeta_\parallel>0$ and $\zeta_\perp>-\mu$.

When the signs of the active coefficients are such that the columnar phase is stable, the equal-time correlators of the displacement fluctuations are 
\begin{equation}
	\langle u_l({\bf q},t)u_l({\bf q}',t)\rangle=\frac{T\delta({\bf q}+{\bf q}')}{q^2[(2\mu+\lambda+\zeta_\perp)\sin^2\theta+\zeta_\parallel\cos^2\theta]},
\end{equation}
and
\begin{equation}
	\langle u_t({\bf q},t)u_t({\bf q}',t)\rangle=\frac{T\delta({\bf q}+{\bf q}')}{q^2[(\mu+\zeta_\perp)\sin^2\theta+\zeta_\parallel\cos^2\theta]}.
\end{equation}
The displacement fluctuations thus diverge as $1/q^2$ along \emph{all} directions of the wavevector space unlike in passive columnar liquid crystals (i.e. in the absence of $\zeta$s) where displacement fluctuations for wavevectors with $\theta=0$ diverges as $1/q^4$. This further implies that the linear roughness exponent for displacement fluctuations is $\chi=-1/2$ (for a three-dimensional columnar phase). Since the fluctuations scale as $1/q^2$ for all directions of the wavevector space, the linear anisotropy exponent is $1$. Finally, since the relaxation rate of displacement fluctuations is $\mathcal{O}(q^0)$ (as a result of Stokesian hydrodynamics), the linear dynamical exponent is $z=0$. This last implies that the hydrodynamic behaviour of active columnar systems in three dimensions are completely described by the linear theory and the linear exponents discussed here are \emph{exact}. That is, there is no relevant nonlinearity. To see this, note that any nonlinear term must at least be of the order $\mathcal{O}(q^0u^2)$ which would have a scaling dimension of $z+\chi$. Since, within the linear theory, $z=0$ and $\chi<0$, all such nonlinear terms would be irrelevant. Furthermore, active \emph{suppression} of displacement fluctuations which render all nonlinearities irrelevant
also cuts-off the low-frequency viscosity divergences known to affect \emph{passive} columnar liquid crystals \cite{ramaswamy1983breakdown}; these now go to a constant value at small frequencies. Since the displacement fluctuations in active columnar and lamellar materials have the same low-frequency, small wavenumber scaling, viscosities are non-divergent at small frequencies even in those systems. 

\subsection{Chiral apolar active columnars}
\label{act_apol}
Chirality couples longitudinal and transverse displacement fluctuations at $\mathcal{O}(q)$:
\begin{multline}
	\partial_t u_l=-\cos^2\theta\left[\frac{2\mu+\lambda}{\eta}\sin^2\theta+\frac{\zeta_\perp}{\eta}\sin^2\theta+\frac{\zeta_\parallel}{\eta}\cos^2\theta\right]u_l\\-\frac{iq\cos\theta}{\eta}\left[\zeta^c_1\cos^2\theta(1+\sin^2\theta)+\zeta^c_4\sin^2\theta(1+\cos^2\theta)-(\zeta^c_2+\zeta^c_3)\cos^2\theta\sin^2\theta\right]u_t
	+\xi_l
\end{multline}
and
\begin{equation}
	\partial_t u_t=-\left[\frac{\mu}{\eta}\sin^2\theta+\frac{\zeta_\perp}{\eta}\sin^2\theta+\frac{\zeta_\parallel}{\eta}\cos^2\theta\right]u_t+\frac{iq\cos\theta}{\eta}(\zeta^c_1\cos^2\theta+\zeta^c_2\sin^2\theta)u_l
	+
	\xi_t.
\end{equation}
These expressions are fairly complicated and difficult to read. However, taking a one activity constant approximation, the dynamics become significantly simpler.
\begin{equation}
	\partial_t u_l=-\cos^2\theta\left[\frac{2\mu+\lambda}{\eta}\sin^2\theta+\frac{\zeta}{\eta}\right]u_l-\frac{i\zeta_cq\cos\theta}{\eta}u_t+\xi_l
\end{equation}
and
\begin{equation}
	\partial_t u_t=-\left[\frac{\mu}{\eta}\sin^2\theta+\frac{\zeta}{\eta}\right]u_t+\frac{iq\zeta_c\cos\theta}{\eta}u_l
	+\xi_t.
\end{equation}
The chiral terms couple the longitudinal and transverse displacement fluctuations. Note that, despite appearances, the $\zeta_c$ term here is not like a Poisson-bracket (i.e. reactive) coupling but a dissipative one. The eigenfrequencies are
\begin{equation}
	\omega=-\frac{i}{2\eta}\left[[\zeta(1+\cos^2\theta)+\sin^2\theta\{(2\mu+\lambda)\cos^2\theta+\mu\}]\pm\sqrt{[\zeta(1-\cos^2\theta)-\sin^2\theta\{(2\mu+\lambda)\cos^2\theta-\mu\}]^2+4\zeta_{c}^2q^2\cos^2\theta}\right].
\end{equation}
The effect of the chiral active stress is most prominent for a wavevector purely along the column direction i.e., when $\theta=0$. In this case, the eigenfrequency becomes
\begin{equation}
	\omega=-\frac{i}{\eta}[\zeta\pm\zeta_c q].
\end{equation}
That is, chirality yields higher order in wavenumber corrections to growth and decay rates of displacement fluctuations.

\subsection{Chiral and polar active columnars}
\label{act_pol}
Using the dynamical equations obtained in Sec. \ref{polcolsec}, we get the linearised equations for the dynamics of longitudinal and transverse displacement fields of chiral and polar active columnar liquid crystals:
\begin{equation}
	\partial_t u_l=-\cos^2\theta\left[\frac{2\mu+\lambda}{\eta}\sin^2\theta+\frac{\zeta_\perp}{\eta}\sin^2\theta+\frac{\zeta_\parallel}{\eta}\cos^2\theta\right]u_l-\cos^2\theta\frac{\zeta_{pc}\sin^2\theta}{\eta}u_t+\xi_l
\end{equation}
and
\begin{equation}
	\partial_t u_t=-\left[\frac{\mu}{\eta}\sin^2\theta+\frac{\zeta_\perp}{\eta}\sin^2\theta+\frac{\zeta_\parallel}{\eta}\cos^2\theta\right]u_t+\frac{\zeta_{pc}\sin^2\theta}{\eta}u_l+\xi_t.
\end{equation}
Again, the chiral polar term couples longitudinal and transverse fluctuations. Notice that $\zeta_{pc}$ looks almost like a \emph{reactive} Onsager coupling (except for the $\cos^2\theta$ required by incompressibility in the $u_l$ equation) between $u_l$ and $u_t$. This excitatory-inhibitory coupling in \emph{non-reciprocal}: Poisson bracket or antisymmetric couplings relate quantities that have \emph{opposite} signs under time-reversal in detailed balance obeying dynamics. Here, they relate two {displacement} fields, which obviously have the same sign under time-reversal. 

The odd-elastic coupling vanishes for fluctuations transverse to the crystalline directions. While, as we discussed in Sec. \ref{polcolsec}, the odd elastic stress in polar columnar materials is analogous to that in two-dimensional chiral active solids, we will now demonstrate that the mode structure of polar columnar liquid crystals is very different from a two-dimensional active solid. This is because the columnar is a \emph{three-dimensional} material with a three-dimensional incompressibility constraint.
In two-dimensional poroelastic solids (with or without odd elasticity), the relaxation rate of $u_l$ fluctuations scale as $\sim q^2$ for all wavevector directions due to the two-dimensional incompressibility constraint (these would appear from permeative dynamics of poroelastic solid). Further, any coupling to $u_t$ would also appear at $\mathcal{O}(q^2)$. 

In contrast, in columnar materials, for wavevectors which have components both in the crystalline plane and along the $\hat{{\bf z}}$ directions, i.e., for $\theta\neq 0$ and $\theta\neq \pi/2$, odd-elasticity leads to a wavenumber-independent excitatory-inhibitory coupling between $u_l$ and $u_t$. This leads to an optical oscillatory mode in this overdamped system. 
Taking $\zeta_\perp=\zeta_\parallel=\zeta$ for simplicity, we get the eigenfrequencies
\begin{equation}
	\omega=-\frac{i}{2\eta}\left[[\zeta(1+\cos^2\theta)+\sin^2\theta\{(2\mu+\lambda)\cos^2\theta+\mu\}]\pm\sin^2\theta\sqrt{[\zeta-\{(2\mu+\lambda)\cos^2\theta-\mu\}]^2-4\zeta_{pc}^2\cos^2\theta}\right].
\end{equation}
The eigenfrequencies are wavenumber-independent because of the overdapmed Stokesian hydrodynamics, yet, as discussed in the main text, they can have a real part for large enough $\zeta_{pc}$. Indeed, such a real part exists as long as 
\begin{equation}
	|\zeta-(2\mu+\lambda)\cos^2\theta+\mu|<2|\zeta_{pc}\cos\theta|
\end{equation}
for \emph{some} $\theta\neq 0$. For instance, for a perturbation with a wavevector making an angle $\pi/4$ with $\hat{{\bf z}}$, the eigenfrequency is 
\begin{equation}
	\omega(\theta=\pi/4)=-\frac{i}{8\eta}\left[6\zeta+4\mu+\lambda\pm\sqrt{(\lambda-2\zeta)^2-8\zeta_{pc}^2}\right],
\end{equation}
which leads to oscillations when $\zeta_{pc}>|2\zeta-\lambda|/2\sqrt{2}$.

\section{Density fluctuations in a columnar system}
\label{densfluc}
We have, till now, been silent about the fluctuations of the density of the active species $\delta c=c-c_0$ about a mean value $c_0$ in columnar systems (the \emph{joint} density of the active species and the fluid is also conserved and accounted for by the incompressibility constraint). We will now demonstrate that like all translation symmetry-broken systems -- such as smectics or solids -- the static structure factor of density fluctuations $\langle|\delta c(q, t)|^2\rangle$ goes to a finite value as $q\to 0$. The displacement field equation has a coupling to the density field due to a free-energetic term $w\delta c\nabla_\perp\cdot{\bf u}_\perp$ (a free energetic coupling between the polarisation and density $\propto \delta c\nabla\cdot{\bf P}$ can be discounted since it is higher order in gradients compared to this; since $\delta{\bf P}_\perp\sim\partial_z{\bf u}_\perp$, this term is $\propto\delta c\partial_z\nabla_\perp\cdot{\bf u}_\perp$). However, a lower order in wavenumber coupling to the density field arises from the active flows. The force balance equation for \emph{all} columnar systems contains gradients of density. This arises from multiple sources: i. There is an active stress $\propto g(c)\hat{n}\hat{n}$ in apolar columnar systems or $\propto g(c)\hat{P}\hat{P}$ in polar columnar system. This leads to a linearised, force density $\propto\partial_z\delta c\hat{{\bf z}}$. ii. A force density $\propto\nabla_\perp\delta c$ appears both from the free energy coupling between density and the displacement fields and from active forces. However, a term $\propto\nabla\delta c$ can be absorbed into the pressure which, in incompressible fluids, is just a Lagrange multiplier that enforces the incompressibility constraint. Therefore, it is enough to retain either the $\partial_z\delta c\hat{{\bf z}}$ or the $\nabla_\perp\delta c$ force density; we choose to retain the former. A force density $\zeta_c\partial_z\delta c$ leads to the in-plane velocity field (in Fourier space)
\begin{equation}
	{\bf v}_\perp=-i\zeta_c\frac{q_z^2}{\eta q^4}q_\perp\delta c,
\end{equation}
which implies that the $u_t$ dynamics doesn't couple to the density field while the $u_l$ dynamics is augmented by a term $-i\zeta_c\cos^2\theta|\sin\theta|\delta c/q$. 

In polar columnars, there is a mass current ${\bf J}_P=\tilde{g}(c){\bf P}$ which accounts for the propulsion of the columns. This implies that the density equation becomes
\begin{equation}
	\partial_t\delta c=-g'\partial_z\delta c-g_0\partial_z\nabla_\perp\cdot{\bf u}_\perp+D_c\nabla^2\delta c+\xi_c=-ig'q\cos\theta\delta c+g_0q^2|\sin\theta|\cos\theta u_l-D_c q^2\delta c+\xi_c,
\end{equation}
where $g'=\partial_c g(c)|_{c=c_0}$, $g_0=g(c_0)$ and $\xi_c$ is a conserving noise.
All other couplings to the displacement field appear at higher orders in gradients. These include the currents $\partial_z\nabla_\perp\cdot{\bf u}_\perp\hat{{\bf z}}$, $\partial_z^2{\bf u}_\perp$, $\nabla_\perp\nabla_\perp\cdot{\bf u}_\perp$. In chiral materials, there is also a current $\propto \partial_z(\nabla_\perp\times{\bf u}_\perp)$. The displacement field dynamics for polar columnars is
\begin{equation}
	\label{ulwithrho}
	\partial_t u_l=-\cos^2\theta\left[\frac{2\mu+\lambda}{\eta}\sin^2\theta+\frac{\zeta}{\eta}\right]u_l-\cos^2\theta\frac{\zeta_{pc}\sin^2\theta}{\eta}u_t-i\zeta_c\frac{\cos^2\theta|\sin\theta|}{\eta q}\delta c+\xi_l
\end{equation}
and
\begin{equation}
	\partial_t u_t=-\left[\frac{\mu}{\eta}\sin^2\theta+\frac{\zeta}{\eta}\right]u_t+\frac{\zeta_{pc}\sin^2\theta}{\eta}u_l+\xi_t.
\end{equation}
The density correlator is particularly simple to calculate in the limit $\zeta_{pc}\to 0$ i.e., for a polar but achiral columnar system. The scaling of the correlator remains unchanged even for $\zeta_{pc}\neq 0$. For $\zeta_{pc}=0$, the $u_t$ dynamics decouple and the coupled $c$ and $u_l$ dynamics have the eigenfrequencies 
\begin{multline}
	\omega_\pm=-\frac{i}{2}\Bigg[\cos^2\theta\left(\frac{2\mu+\lambda}{\eta}\sin^2\theta+\frac{\zeta}{\eta}\right)+Dq^2+ig'q\cos\theta\\\pm\sqrt{\left\{\cos^2\theta\left(\frac{2\mu+\lambda}{\eta}\sin^2\theta+\frac{\zeta}{\eta}\right)-Dq^2-ig'q\cos\theta\right\}^2-4i\frac{\zeta_c}{\eta} g_0q\sin^2\theta\cos^3\theta}\Bigg].
\end{multline}
These can be written in the form $\omega_\pm=c_\pm q\cos\theta-i\Xi_\pm$, where $\Xi_+\sim \mathcal{O}(q^0)$ and $\Xi_-\sim \mathcal{O}(q^2)$ . With this, the density correlator becomes
\begin{equation}
	\langle|\delta c({\bf q},\omega)|^2\rangle\approx\frac{\langle\xi_l({\bf q},\omega)\xi_l(-{\bf q},-\omega)\rangle g_0^2q^4\sin^2\theta\cos^2\theta}{[(\omega-c_+q\cos\theta)^2+\Xi_+^2][(\omega-c_-q\cos\theta)^2+\Xi_-^2]}=\frac{2T g_0^2q^2\sin^2\theta\cos^4\theta}{\eta[(\omega-c_+q\cos\theta)^2+\Xi_+^2][(\omega-c_-q\cos\theta)^2+\Xi_-^2]}
\end{equation}
Therefore, the equal-time density correlator becomes 
\begin{equation}
	\langle|\delta c({\bf q},t)|^2\rangle=\frac{T g_0^2q^2\sin^2\theta\cos^4\theta|\Xi_++\Xi_-|}{\eta\Xi_+\Xi_-[(c_+-c_-)q^2\cos^2\theta+(\Xi_++\Xi_-)^2]}.
\end{equation}
Since $\Xi_+\sim q^0$ and $\Xi_-\sim q^2$, $\Xi_+\Xi_-\sim q^2$ and $\Xi_++\Xi_-\sim q^0$. Therefore, as $q\to 0$, $\langle|\delta c({\bf q},t)|^2\rangle$ goes to a $q$-independent constant value. While the coefficient of the $q^0$ part of $\Xi_+$ vanishes in some directions, specifically for $\cos\theta=0$, the density and displacement fields decouple along those directions to lowest order in wavenumbers. It is easy to see that even in those directions $\langle|\delta c({\bf q},t)|^2\rangle\sim q^0$ using the conserving noise in the density equation.

The density correlator goes to a constant at small wavenumbers in apolar systems as well (irrespective of whether it is chiral or not). In this case, the $u_l$ equation is the same as \eqref{ulwithrho} with $\zeta_{pc}=0$ and the density equation, to lowest order in gradients, is
\begin{equation}
	\partial_t\delta c=-\gamma_1\partial_z^2\nabla_\perp\cdot{\bf u}_\perp-\gamma_2\nabla_\perp^2\nabla_\perp\cdot{\bf u}_\perp+D_c\nabla^2\delta c+\xi_c=i q^3(\gamma_1 \cos^2\theta+\gamma_2\sin^2\theta)\sin\theta u_l-D_c q^2\delta c+\xi_c,
\end{equation}
where we have ignored the chiral current which appears at the same order in wavenumbers and doesn't change the scaling of the static structure factor of density fluctuations at small wavenumbers. Further, both the eigenvalues are now purely relaxational $\omega_\pm=-i\Xi_\pm$ with $\Xi_+\sim q^0$ and $\Xi_-\sim q^2$. Therefore, $\langle|\delta c({\bf q},t)|^2\rangle\sim q^0$ again at small wavenumbers. In fact, here the couplings $\gamma_i$ to the displacement field don't affect the small wavenumber density structure factor.

\bibliography{ref}

\end{document}